\documentclass[aps, pra, nofootinbib, onecolumn, 11pt, tightenlines,notitlepage, superscriptaddress, longbibliography]{revtex4-1}
\usepackage[letterpaper, total={6in, 9in}]{geometry}
\usepackage{amssymb,mathtools}
\usepackage{amsmath,dsfont,physics}
\usepackage{mathrsfs}
\usepackage{graphicx}
\usepackage{dcolumn}
\usepackage[colorlinks=true,allcolors=green]{hyperref}
\usepackage{verbatim}
\usepackage{natbib}
\usepackage{cleveref}
\usepackage{float}
\usepackage{color,xcolor}
\usepackage[utf8]{inputenc}
\usepackage[english]{babel}
\usepackage{amsthm}
\usepackage{appendix}
\usepackage{pgfplots,tikzscale}
\usepackage{soul}
\pgfplotsset{compat=1.5}

\newcommand{\dket}[1]{\lvert #1 \rangle\!\rangle}
\newcommand{\dbra}[1]{\langle\!\langle #1 \rvert}
\newcommand{\dbraket}[2]{\langle\!\langle #1 \vert #2 \rangle\!\rangle}

\def \mc #1{\mathcal{#1}}

\setul{0.5ex}{0.3ex}
\setulcolor{red}

\begin{document}

\title{On the freedom in representing quantum operations}
\author{Junan Lin}
\email{j242lin@uwaterloo.ca}
\affiliation{Institute for Quantum Computing and Department of Physics and Astronomy, University of Waterloo, Waterloo, Ontario N2L 3G1, Canada}

\author{Brandon Buonacorsi}
\email{bbuonaco@uwaterloo.ca}
\affiliation{Institute for Quantum Computing and Department of Physics and Astronomy, University of Waterloo, Waterloo, Ontario N2L 3G1, Canada}
\author{Raymond Laflamme}
\affiliation{Institute for Quantum Computing and Department of Physics and Astronomy, University of Waterloo, Waterloo, Ontario N2L 3G1, Canada}
\affiliation{Perimeter Institute for Theoretical Physics, Waterloo N2L 2Y5 ON, Canada}
\author{Joel J. Wallman}
\email{jwallman@uwaterloo.ca}
\affiliation{Institute for Quantum Computing and Department of Applied Mathematics, University of Waterloo, Waterloo, Ontario N2L 3G1, Canada}
\date{\today}

\begin{abstract}
We discuss the effects of a gauge freedom in representing quantum information processing devices, and its implications for characterizing these devices.
We demonstrate with experimentally relevant examples that there exists equally valid descriptions of the same experiment which distribute errors differently among objects in a gate-set, leading to different error rates.
Consequently, it can be misleading to attach a concrete operational meaning to figures of merit for individual gate-set elements.
We propose an alternative operational figure of merit for a gate-set, the mean variation error, and a protocol for measuring this figure.
\end{abstract}

\maketitle

Knowing how to characterize one's control over a quantum system is of utmost importance in quantum information processing.  
An experimentalist requires protocols and metrics that appropriately describe the error rate of their quantum processor.
Quantum mechanics allows us to assign representations to describe the state of quantum objects and processes, and many figures of merit have been developed to evaluate them based on their representation \cite{gilchrist2005distance}.
For example, the fidelity and trace distance are two commonly quoted measures.

Conventional quantum tomography unrealistically assumes that the states and/or measurements being used to probe the unknown operation are ideal.
Recently, gate-set tomography (GST) has been developed to avoid making such assumptions by self-consistently inferring all gate-set elements from experimentally estimated probabilities \cite{Merkel2013,blume2013robust}.
Relaxing these assumptions results in a non-unique representation of the gate-set due to a gauge freedom \cite{blume2013robust,blume2017demonstration}.
Many conventional measures depend on the particular representation for quantum operations.
Therefore, assessing the quality of a quantum device in terms of these metrics applied to non-unique representations may be inaccurate.
Despite the broad conceptual importance of representing quantum operations, the impact of the gauge freedom has only occasionally been analyzed, and primarily in the context of gate-dependent noise in randomized benchmarking \cite{proctor2017randomized,rudnicki2017gauge,wallman2018randomized}.

In this work we clarify how this gauge freedom affects experimental descriptions and demonstrate some of its implications for interpreting experimentally reconstructed representations of quantum objects.
In \cref{exampleSection} we demonstrate with an experimentally motivated example that the gauge freedom makes assigning errors to individual operations ambiguous and demonstrates that the gauge freedom is a separate issue from gate-dependent noise.
In \cref{gaugeSection} we give definitions for gauges and gauge transformations, as well as their role in representing quantum operations as mathematical objects.
In \cref{implicationSection} we discuss the implications brought by this gauge freedom, in particular addressing why many figures of merit such as the diamond norm distance between a measured gate and a target do not have a concrete operational meaning. 
We also mention some common practices in tomography that are related to this problem.
Lastly, in \cref{newMeasureSection}, we define and motivate the mean variation error (MVE), a gauge-invariant figure of merit for gate-sets.
We provide a protocol to experimentally measure the MVE and demonstrate its behaviour relative to randomized benchmarking through numerical simulations.

\section{Assigning errors to operations}\label{exampleSection}
We now illustrate the gauge freedom with a simple, experimentally relevant example, namely, amplitude damping.
A gate-set is a mathematical description of the possible actions executable in an experiment, typically consisting of models for initial states ($\mathds{S}$), gate operations ($\mathds{G}$), and measurements ($\mathds{M}$).
If an experimentalist with an ideal quantum system could initialize a qubit in the state $\ket{0}$, apply an arbitrary unitary gate, and measure the expectation value of $Z$, then their control can be represented by the gate-set
\begin{equation}
\Phi = \left\{ \mathds{S}_\Phi = \begin{pmatrix}
1 & 0\\ 0 & 0
\end{pmatrix},\ \mathds{G}_\Phi = \rm{SU}(2) ,\ \mathds{M}_\Phi = Z\right\}.
\end{equation}
Now suppose that the experimentalist prepares a mixed initial $Z$ state with polarization $\epsilon_1$ and performs a measurement with signal-to-noise ratio $\epsilon_2$.
Suppose further that before each gate is applied, the system undergoes amplitude damping with strength $\gamma$ but that the target Hamiltonian is still implemented perfectly.
The Pauli-Liouville representation (see \cref{appendix} or e.g., \cite{greenbaum2015introduction}) of the noisy gate-set is then
\begin{equation}\label{gateSetPauli}
\Theta = \left\{\mathds{S}_\Theta = \frac{1}{\sqrt{2}} \begin{pmatrix}
1 \\ 0 \\ 0 \\ \epsilon_1
\end{pmatrix},\ 
\mathds{G}_\Theta = \{\mc{UA}_\gamma: U\in\rm{SU}(2)\},\ 
\mathds{M}_\Theta = \sqrt{2} \begin{pmatrix}
0 & 0 & 0 & \epsilon_2
\end{pmatrix} \right\}
\end{equation}
with $\sqrt{2}$ and $\frac{1}{\sqrt{2}}$ being normalization factors, and $\mathcal{U} = U\rho U^\dagger$ denotes the unitary channel acting via conjugation and
\begin{align}
\mc{A}_\gamma = \begin{pmatrix}
1 & 0 & 0 & 0\\
0 & \sqrt{1 - \gamma} & 0 & 0\\
0 & 0 & \sqrt{1 - \gamma} & 0\\
\gamma & 0 & 0 & 1 - \gamma
\end{pmatrix}.
\end{align}

The expectation value of an operator $M$ given an input state $\rho$ is the vector inner product between the Pauli-Liouville representations of the state and measurement operators, 
\begin{equation}
\text{prob} = \dbraket{M}{\rho}.
\end{equation}
If $m$ gates $\mathcal{G}_1,\ldots,\mathcal{G}_m\in\mathds{G}$ are applied to the state in chronological order before the measurement takes place, the expectation value becomes
\begin{equation}
\text{prob} = \dbra{M} \mathcal{G}_{m:1}  \dket{\rho}
\end{equation}
where we use the shorthand notation 
\begin{equation}
	\mathcal{G}_{b:a} \coloneqq \begin{cases}
	\mathcal{G}_b \mathcal{G}_{b-1}...\mathcal{G}_a & \text{if}\ b \geq a \\
	\mathcal{I} & \text{otherwise.}
	\end{cases} 
\end{equation} 

The above probabilities are preserved under the family of gate-set transformations
\begin{equation}\label{GaugeEqn}
\dket{\rho} \rightarrow B \dket{\rho},\ \dbra{M} \rightarrow \dbra{M}B^{-1},\ \mathds{G}_\Phi \rightarrow B\mathds{G}_\Phi B^{-1}
\end{equation}
for some invertible matrix $B$.
Because these probabilities are the only experimentally accessible quantities, the same experimental results can be predicted equally well by these two gate-sets.
This is the gauge freedom inherent in mathematically representing quantum experiments, in analogy with concepts in thermodynamics and electromagnetism \cite{jackson2017nonholonomic}, with $B$ being called a gauge transformation matrix.
The analogy arises from the fact that changing the gauge does not result in observable effects in an experiment, just as changing the electromagnetic gauge would not result in any difference in the measurable electric or magnetic fields.

Generally, a gate-set is considered valid if all quantum states can be represented as density matrices, measurements as expectation values of Hermitian operators, and quantum gates as completely-positive, trace-preserving (CPTP) maps as these conditions ensure that probabilities for arbitrary experiments are positive.
Gauge transformations do not generally preserve these \textit{canonical constraints}, although the resulting gate-set is nevertheless an equally valid mathematical description of the same experiment.

We now present a simple, physically motivated, gauge transformation that yields a gate-set that suggests a different physical interpretation of the experimental system.
Applying the gauge transformation matrix
\begin{equation}\label{gauge}
B=\begin{pmatrix}
1 & 0 & 0 & 0\\
0 & q & 0 & 0\\
0 & 0 & q & 0\\
0 & 0 & 0 & q
\end{pmatrix}
\end{equation}
for any $q\in[-1,1]$ to the noisy gate-set in \cref{gateSetPauli} yields the equivalent gate-set
\begin{equation}\label{gateSetPauli2}
\Theta_q = \left\{\mathds{S}_{\Theta_q} = \frac{1}{\sqrt{2}} \begin{pmatrix}
1 \\ 0 \\ 0 \\ q \epsilon_1
\end{pmatrix},\ 
\mathds{G}_{\Theta_q} = \{\mc{U} \mc{A}_{\gamma,q}: U\in\rm{SU}(2)\},\
\mathds{M}_{\Theta_q} = \sqrt{2} \begin{pmatrix}
0 & 0 & 0 & \frac{\epsilon_2}{q}
\end{pmatrix} \right\},
\end{equation}
where
\begin{align}
    \mc{A}_{\gamma,q} = \begin{pmatrix}
1 & 0 & 0 & 0\\
0 & \sqrt{1 - \gamma} & 0 & 0\\
0 & 0 & \sqrt{1 - \gamma} & 0\\
q\gamma & 0 & 0 & 1 - \gamma
\end{pmatrix}
\end{align}
and we have used the fact that $\mc{U}$ commutes with $B$ for any $U\in\rm{SU}(2)$.
The gauge transformation results in equivalent statistics  but suggests a different noise model, namely, relaxation to a mixed state rather than a pure state (corresponding to a different effective temperature).
As long as $|q|\in[|\epsilon_2|,1]$, the states, measurements, and transformations all satisfy the canonical constraints for gate-set elements.

Note that this gauge freedom does not change the average gate fidelity as $\Tr[\mc{A}_{\gamma,q}]$ is independent of $q$~\cite[Eq. 2.5]{Kimmel2014}.
However, the diamond distance from the identity depends on $q$, with $\|\mc{A}_{\gamma,1} - \mc{I}\|_\diamond \approx 2 \|\mc{A}_{\gamma,0} - \mc{I}\|_\diamond $ for $\gamma\in[0,1]$~\cite{Kueng2016}.
Moreover, this example illustrates that noise can be artificially reassigned to different objects, as the state in \cref{gateSetPauli} is closer to pure than the one in \cref{gateSetPauli2}.
Note that the range of gauge transformations is constrained by $\epsilon_2$ and so cannot significantly change the effective temperature for systems with high quality readout.
We could have added larger errors by considering a non-unital (e.g., $\mc{A}_{\gamma,q}B$) or unitary gauge transformation at the cost of making the errors gate-dependent and consequently giving a more complicated example.
We did not do this as our intent is to clarify that the gauge freedom is more than a basis mis-match~\cite{carignan2018randomized} and is distinct from the issue of gate-dependent noise~\cite{proctor2017randomized, wallman2018randomized, carignan2018randomized}.
In particular, we note that the full effect of the gauge freedom for states and measurements is unknown.

\section{Gauge and Representation of Quantum States}\label{gaugeSection}
We have seen that under realistic circumstances, the same experiment can be described by distinct gate-sets that suggest different physical noise models due to a gauge freedom. 
In this section we illustrate how representations of quantum states are related to the concepts of gauges and gauge transformations.
For clarity we focus on the representation for quantum states, but similar arguments can be made about gate and measurement operations.

\begin{figure}[ht]
\centering
\includegraphics[width=0.7
\textwidth]{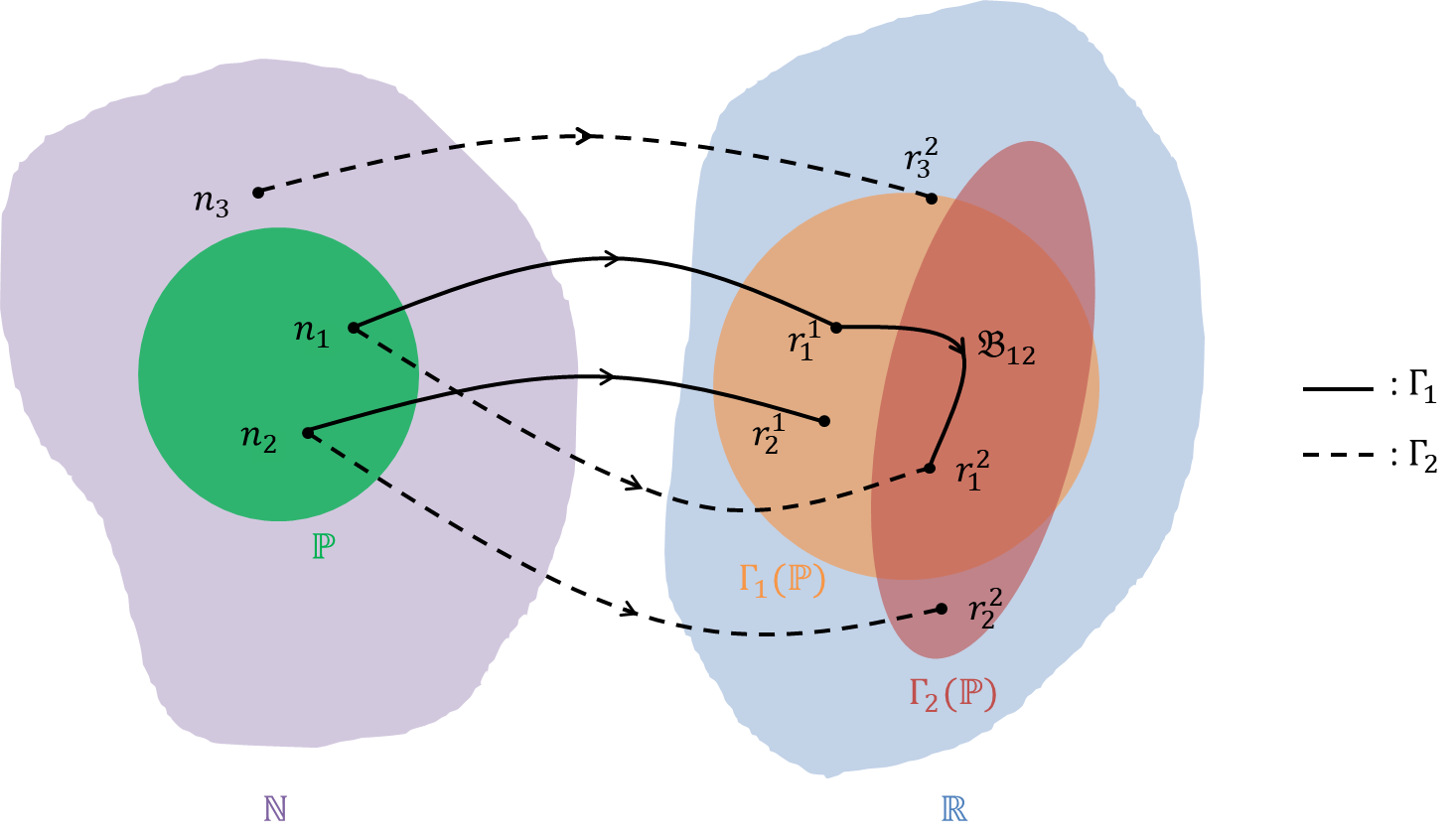}
\caption{Diagram illustrating the distinction between the set of noumenal states $\mathds{N}$ and their mathematical representations $\mathds{R}$. 
For a single qubit, $\mathds{R}$ is the set of $2 \times 2$ matrices.
A gauge is a bijective map $\Gamma:N\to R$.
Let $\mathds{P}$ be the set of all noumenal states that can possibly be prepared physically.
In general, $\mathds{P}$ is unknown, but $\Gamma(\mathds{P})$ is assumed to satisfy the canonical constraints, that is, to be the set of density matrices.
In our example, $\Gamma_1(\mathds{P})$ is the set of density operators in the corresponding Hilbert space and $\mathfrak{B}_{12}$ is a gauge transformation from $\Gamma_1$ to $\Gamma_2$.
Whether an object in $\mathds{R}$ directly corresponds to objects in $\mathds{P}$ or not depends on the particular gauge under which the object is represented.
For example, $r_2^1 = \Gamma_1(n_2)$ is a density operator while $r_2^2 = \Gamma_2(n_2)$ is not, despite both being the image of the same physical state.}  
\label{VennDiag}
\end{figure}

From the point of view of scientific realism, the apparatus (e.g., a qubit) has a physical existence and properties (which may be relative to the environment) independent of our representation.
We describe the abstract state of this physical object as a \textit{noumenal state} following the terminology in \cite{brassard2017equivalence}, denoted as $\mathds{N}$ in Figure \cref{VennDiag}.
Here we slightly change their definition to include in $\mathds{N}$ both physically allowed (denoted as $\mathds{P}$) and forbidden states, such that the set $\mathds{N}$ contains both states that the system can be in, and ones that it cannot be in based on the physics.
Quantum mechanics allows us to assign to each noumenal state a mathematical \textit{representation} which is an element of a Hilbert space $\mathcal{H}^d$: for example, one can associate the system with a matrix that summarizes its properties, and the set of all $d \times d$ matrices is called $\mathds{R}$ in the same figure.
Such an association is what we call a \textit{gauge} $\Gamma$, which is a bijective map from $\mathds{N}$ to $\mathds{R}$: the bijectivity of the map should be clear from our inclusion of physically-forbidden states in $\mathds{N}$, which allows assigning ``some state'' to every $d \times d$ matrix.
Different choices of $\Gamma$ thus correspond to different mathematical descriptions of the noumenal states.

The common formulation of quantum mechanics says that every state of a quantum object can be described by a density operator \cite{hardy2001quantum}, which is a part of the canonical constraints defined in \cref{exampleSection}.
This means that there exists a \textit{canonical gauge} \begin{equation}
\Gamma_1:\ N \rightarrow R,\ \Gamma_1(\mathds{P}) = \mathds{D}_d
\end{equation}
where $\mathds{D}_d$ is the set of $d \times d$ density operators. In fact, there exists a family of canonical gauges that are all related to $\Gamma_1$ through unitary gauge transformations which preserves the shape of $\mathds{D}_d$.
Satisfying the canonical constraint implies that we should work in one of these canonical gauges.
Now, consider another gauge $\Gamma_2$ which can be converted from $\Gamma_1$ with a \textit{gauge transformation} $\mathfrak{B}_{12}$, defined by
\begin{equation}
\mathfrak{B}_{12} \coloneqq \Gamma_2 (\Gamma_1^{-1}),\ \mathfrak{B}_{12}(r_{*}^{1}) = r_*^2
\end{equation}
where the $r$'s are members in $\mathds{R}$ and the superscript denotes the gauge in which they are represented.
In the light of \cref{GaugeEqn}, this transformation can be represented in Pauli-Liouville representation as
\begin{equation}\label{gauge B12}
\dket{\mathfrak{B}_{12} (\rho)} \coloneqq B_{12} \dket{\rho},\ \dbra{\mathfrak{B}_{12}(M)} \coloneqq \dbra{M}B_{12}^{-1},\ \mathds{G}_{\mathfrak{B}_{12}(\Phi)} \coloneqq  B_{12}\mathds{G}_{\Phi} B_{12}^{-1} 
\end{equation}
As a subset of $\mathds{R}$, $\Gamma_1(\mathds{P})$ is generally not invariant under an arbitrary gauge transformation: consider a general trace-preserving transformation given by the following transformation matrix
\begin{equation}
B_{12} = \begin{pmatrix}
1 & 0\\ \vec{x} & y
\end{pmatrix}
\end{equation}
where $\vec{x}$ is a $(d^2-1) $ by $ 1$ real vector and $y$ is a $(d^2-1) $ by $ (d^2-1)$ real matrix: the image of this affine transformation of $\Gamma_1(\mathds{P})$ is a different subset of $\mathds{R}$. 
Such a gauge is perfectly valid in principle, provided that \textit{all} the gates and measurement operators are transformed according to \cref{gauge B12} as well, even though $\Gamma_2(\mathds{P})$ is no longer the set of density operators. 

The existence of a non-canonical gauge implies, for example, that a physical state may or may not be represented by a density operator: as illustrated in \cref{VennDiag}, $r_1^2 \in \Gamma_1(\mathds{P})$ whereas $r_2^2 \notin \Gamma_1(\mathds{P})$.
Similarly, a density operator in a non-canonical gauge does not necessarily correspond to a physical state, as $r_3^2 \in \Gamma_1(\mathds{P})$ but $n_3 \notin \mathds{P}$.
One example for the state $n_3$ is a qubit state represented as $\frac{1}{2} (I+\frac{10}{9} \sigma_z)$ in a canonical gauge. 
It is not positive semidefinite and thus lies outside $I_1$, representing an abstract state the qubit cannot be in. 
Now, using $B_{12} = B$ from \cref{gauge} with $q = \frac{9}{10}$, the image of $n_3$ under $\Gamma_2$ becomes $\frac{1}{2} (I+ \sigma_z)$, which \textit{is} a density operator, but only as a consequence of this non-canonical gauge.
We conclude that if the gauge is unknown, the mathematical representation does not imply the noumenal state is physically possible.
Representations satisfying the canonical constraints are easier to work with, so it is often implicitly assumed that all gate-set elements (obtained from a tomography experiment, for example) are expressed in a canonical gauge. 
However, this assumption can only be verified by performing perfect experiments, which are axiomatically the operations specified by the canonical constraints (up to a unitary change of basis).

\section{Operational interpretations of figures of merit}\label{implicationSection}
The existence of this gauge freedom has direct implications for figures of merit used to evaluate quantum operations.
The main problem is that there is no way to know whether an experimentally-determined gate-set element is expressed in a canonical gauge.
We have already seen in \cref{exampleSection} that by changing the gauge, the states can appear as having different expressions; the same holds true for gates and measurement operators.

From quantum information theory, we have successfully attached some operational meanings to various distance metrics: an important example is the interpretation for the diamond norm distance between two channels $\mc{A}$ and $\mc{B}$ as the maximum distinguishability between output states under a fixed input~\cite{watrous2018theory}.
Mathematically,
\begin{equation}\label{optimalInputOutput}
\frac{1}{2} \norm{\mc{A} - \mc{B}}_\diamond =  \max_{M\in\Gamma(\mathds{M}), \rho\in\Gamma(\mathds{P})} \dbra{M} (\mc{A} - \mc{B}) \otimes I) \dket{\rho}
\end{equation}
where $\mathds{P}$ and $\mathds{M}$ are the set of physically possible states and measurements respectively. 
This operational meaning is gauge invariant, provided one consistently transforms $\mc{A}$, $\mc{B}$, $\Gamma(\mathds{P})$, and $\Gamma(\mathds{M})$.
However, when $\mc{A}$ is an experimentally reconstructed gate and $\mc{B}$ is its ideal target, $\Gamma(\mathds{P})$ and $\Gamma(\mathds{M})$ are unknown and so the above maximization that leads to its operational meaning cannot be performed.
To obtain concrete numbers, people calculate
\begin{equation}
\frac{1}{2} \norm{\mc{A} - \mc{B}}_\diamond =  \max_{M\in \mu(\mathds{D}_{d}), \rho\in\mathds{D}_{d}} \dbra{M} (\mc{A} - \mc{B}) \otimes I) \dket{\rho}
\end{equation}
where $\mu(\mathds{D}_{d})$ is the set of all POVMs.
However, this assumes that the reconstructed $\mc{A}$ and the ideal target $\mc{B}$ are expressed in a canonical gauge.
While $\mc{B}$ is an ideal gate, its representation may not be unitary in a non-canonical (and unknown) gauge.
Other works have reported that the quantity $\frac{1}{2} \norm{\mc{A} - \mc{B}}_\diamond$ can be changed by changing the gauge and used this to minimize reported error rates~\cite{blume2017demonstration}, however, such changes are obtained by implicitly changing the set of physically allowed states and measurements.
Note that even in one special case of interest where $\mc{B} = \mc{I}$, which is gauge invariant, $\mc{A}$ is still reconstructed in an unknown gauge.

We briefly discuss several common practices related to this gauge freedom in quantum tomography.
First, a common statistical method used in tomographic reconstructions is Maximum Likelihood Estimation (MLE), which takes the  estimated gate-set to be the one that maximizes the likelihood function of obtaining the experimental data, while restricting the gate-set elements to satisfy the canonical physicality constraints \cite{medford2013self,brida2012quantum}.
However, all gauge-equivalent gate-sets are equally likely to produce the data by definition.
In the process of optimization, one will find that the likelihood function profile has the same value wherever two points are related by a gauge-transformation, and the actual output is largely a matter of the optimization algorithm and the initial parameters \cite{blume2015turbocharging}.

Second, the process known as ``gauge optimization'' is commonly adopted in GST experiments whereby the gauge transformation matrix $B$ is varied to minimize the distance from the target gate-set according to a (non-gauge-invariant) weighted distance measure \cite{blume2017demonstration}.
Such optimization undermines a common use of tomography, namely assessing the performance of a system against some external threshold (e.g., a fault-tolerance threshold) because this optimized gauge is just as arbitrary as any other gauge.
Specifically, one still cannot know whether the resultant gate-set is a faithful representation of the apparatus, in particular, whether the states and measurements that satisfy the canonical constraints are actually the images of the set of physically possible states and measurements respectively.

Furthermore, the optimization can change the relative size of errors from different components of the device, leading to misunderstandings about their relative quality.
This misidentification of error was recently observed experimentally in a trapped ion processor: in particular, fig. 4(h) in~\cite{Mavadia2018} demonstrates that the gauge optimization procedure may effectively cancel some gate errors that were purposely added, resulting in a smaller ``diamond norm distance'' than expected, which implies unrealistically good quantum gates.
In that paper, and in other systems where, for example, a basis change in the classical software is used to achieve certain ``virtual'' gates~\cite{mckay2017efficient,knill2000algorithmic}, experimentalists have \textit{a priori} information about which operations are better.
This is also true for state and measurement operations in some systems: for example, in an NMR spectrometer there is a well-defined $Z$ direction set by the external magnetic field, which defines the initial qubit states with which all other operations are calibrated.
But incorporating such information into the optimization procedure is a challenging task, and the resulting representation will only be as reliable as the prior information.

\section{A Gauge-Invariant Measure for Gate-Sets}\label{newMeasureSection}
The gauge freedom prevents one from using conventional distance measures to faithfully evaluate the quality of individual quantum operations.
Note that our discussion is carried out in the absence of any additional errors such as finite-counting, and in a real experiment the situation becomes even more complicated.
Fundamentally, this problem is due to the limited information that can be gained from experimental probabilities.
A gauge-transformation re-assigns state, gate, and measurement ``errors'' by adjusting their relative appearance in different representations, while keeping the experimental measurables unchanged, although some degrees of freedom can be fixed by convention (e.g., that the state preparation is diagonal in the $Z$ eigenbasis).

We now propose a gauge-invariant figure of merit for a \textit{gate-set}.
As far as we know, this is the first fully gauge-invariant measure, addressing a problem raised in Ref.~\cite{blume2015turbocharging}.
Let $\Phi$ denote the gate-set $\{\mathds{S}, \mathds{G}, \mathds{M}\}$ and $C$ denote a particular experiment with input state $\rho \in \mathds{S}$, measurement $M \in \mathds{M}$, and a set of $m$ gates $\mc{G}_1...\mc{G}_m$ each selected from $\mathds{G}$.
The only observable property of an experiment $C$ is the probability distribution over outcomes.
We can quantify the error of the experiment by the total variation distance between the observed and ideal distributions over outcomes,
\begin{equation}\label{delta-d defn}
\delta d (C,\tilde{C}) \coloneqq \frac{1}{2} \sum_i \abs{\Tr[\tilde{M}_i^\dagger \tilde{G}_{m:1} (\tilde{\rho})] - \Tr[M_i^\dagger G_{m:1} (\rho)]} 
\end{equation}
where the tilde represents real versions of the operations, and the subscript $i$ denotes different measurement operators such that $\sum_i M_i = I$.
For a particular experiment, this number only depends on outcome probabilities, and therefore is gauge-invariant by definition.
The total variation distance is a metric between two probability distributions (for further motivation and background on this metric, see chapter 4.1 of~\cite{levin2017markov}).
Denoting the set of all experiments with $m$ gates by $\mathds{A}_{m}$, we further define the \textit{Mean Variation Error} (MVE) over $\mathds{A}_{m}$ with the underlying gate-set $\Phi$ as
\begin{equation}\label{MVE definition}
x(\Phi, m) \coloneqq \frac{1}{|\mathds{A}_{m}|}\sum_{C\in\mathds{A}_m} [\delta d (C, \tilde{C})]
\end{equation}
Note that MVE is a special case of the \textit{mean variation distance} (MVD) between two gate-sets $\Phi_1$ and $\Phi_2$ with the same number of elements in $\mathds{S}$, $\mathds{G}$ and $\mathds{M}$, where $\Phi_2$ is chosen to be an ideal version of $\Phi_1$.
The MVD is a metric between two gate-sets, up to a gauge transformation, if the images of state preparation and measurement processes are tomographically complete.
Also note that the size of $\mathds{A}_{m}$ is given by $\abs{\mathds{A}_{m}} = \abs{\mathds{S}} \abs{\mathds{G}}^m \abs{\mathds{M}}$.

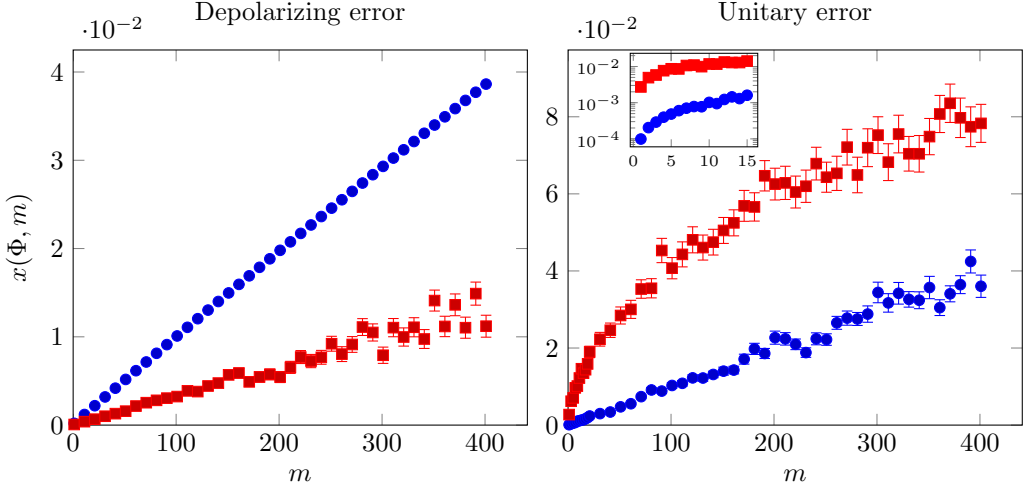
\begin{figure}[ht]
\centering
\begin{tikzpicture}
\begin{axis}[xmin=0,ymin=0,width = 0.5\linewidth,xlabel={$m$},scaled y ticks=base 10:2 
/pgf/number format/sci subscript, ylabel={$x(\Phi,m)$}, title={Depolarizing error}]
\addplot+[error bars/.cd,y dir=both,y explicit]
table [x=m, y=mean, y error=std, only marks, col sep=comma] {depolarizing_identity.csv};
\addplot+[error bars/.cd,y dir=both,y explicit]
table [x=m, y=mean, y error=std, only marks, col sep=comma] {depolarizing_general.csv};
\end{axis}
\end{tikzpicture}
\begin{tikzpicture}
\begin{axis}[xmin=0,ymin=0,width = 0.5\linewidth,xlabel={$m$}, title={Unitary error},scaled y ticks=base 10:2
/pgf/number format/sci subscript]
\addplot+[error bars/.cd,y dir=both,y explicit]
table [x=m, y=mean, y error=std, only marks, col sep=comma] {unitary_identity.csv};
\addplot+[error bars/.cd,y dir=both,y explicit]
table [x=m, y=mean, y error=std, only marks, col sep=comma] {unitary_general.csv};
\end{axis}
\begin{semilogyaxis}[at={(rel axis cs:0,1)},anchor={outer north west},tiny,width = 0.215\linewidth]
\addplot+[error bars/.cd,y dir=both,y explicit]
table [x=m, y=mean, only marks, col sep=comma] {shortunitary_identity.csv};
\addplot+[error bars/.cd,y dir=both,y explicit]
table [x=m, y=mean, only marks, col sep=comma] {shortunitary_general.csv};
\end{semilogyaxis}
\end{tikzpicture}
\caption{Simulated mean variation error from \cref{MVE definition} under two error models for a gate-set with $\rho = M_1 = \ketbra{0}{0}$, and $\mathds{G} = Cl_1 $ being the 1-qubit Clifford group. The depolarizing error channel is $\mathcal{E}_D (\rho,r) = (1-2r)\rho + r I $, whereas the unitary error is $\mathcal{E}_U (\rho,\theta) = e^{-i \theta Z} $ with $\theta = \arccos(\sqrt{1-3r/2})$, such that the error channel on every gate has an averaged infidelity of  $r = 10^{-4}$. Blue circles indicate self-inverting (identity) circuits whereas red squares indicate random circuits. Each point is generated from averaging 200 random circuits with length $m$; error bars are standard error in the mean and data show significant spread for unitary error. MVE may have different behaviors under different error types ($m$ or $\sqrt{m}$) for a random circuit, as compared to the linear behavior for a self-inverting circuit. The inset in the second plot is a zoom-in view for small $m$, showing the significant underestimation of MVE by restricting to self-inverting circuits.}
\label{New-metric-plot}
\end{figure}

The MVE quantifies on average how well the apparatus performs a random experiment from the gate-set.  
In the case where the measurement is a projective measurement in the basis of the initial state (i.e., $\rho$ = $M_i$ for some $i$) and the gate sequence is self-inverting\footnote{A self-inverting circuit is one where all the gates performed in the circuit compose to the identity operation in the absence of gate errors.}, $\delta d(C,\tilde{C})$ can be simplified as
\begin{equation}
\begin{aligned}
\delta d (C, \tilde{C}) &= \frac{1}{2} \left(\abs{\Tr[\tilde{M}_i^\dagger \tilde{G}_{m:1} (\tilde{\rho})] - 1} + \sum_{j \neq i} \abs{\Tr[(I - \tilde{M}_{j})^\dagger \tilde{G}_{m:1} (\tilde{\rho})] - 0} \right)\\
&= 1 - \Tr[\tilde{M}_i^\dagger \tilde{G}_{m:1} (\tilde{\rho})]
\end{aligned}
\end{equation}
whose average over $\mathds{A}_m$ is just 1 minus the ``survival probability'' plotted in a conventional randomized benchmarking experiment.
When $\mathds{G}$ is a unitary 2-design, the MVE restricted to self-inverting gate sequences is well-approximated by a linear relation to first order in the average error rate \cite{wallman2018randomized,proctor2017randomized}.

However, for generic gate sequences, the MVE behaves differently depending on the underlying error model.
This behavior provides additional information about the underlying error mechanism compared to a conventional randomized benchmarking experiment~\cite{wallman2015bounding}.
To illustrate this, we simulated random circuits of varying length $m$ sampled from the gate-set $\Phi = \{\mathds{S} = \ketbra{0}, \mathds{G} = \textit{Cl}_1, \mathds{M} = \ketbra{0}\}$ (with $\textit{Cl}_1$ denoting the set of 1-qubit Clifford gates), where erroneous gates are represented as $\tilde{\mathcal{G}} =\mathcal{E} \mathcal{G}$ for a fixed error channel $\mathcal{E}$.
We simulated two types of random circuits: circuits from the entire set of possible experiments allowed by the gate-set, and circuits restricted to self-inverting gate sequences. 
In both simulations, the state and measurements are assumed to be error-free.
The results are shown in \cref{New-metric-plot}. 
When the error is a depolarizing channel, the MVE scales linearly with the gate sequence length $m$ for both random and self-inverting circuits, with the slope for random circuits being $\sim 1/3$ the slope for self-inverting circuits.  
This is because when the state is transformed onto the xy-plane of the Bloch sphere right before measurement (which happens about 2/3 of the time), the depolarizing channel does not affect the outcome probability of a z-axis measurement, resulting in an MVE of 0 for those circuit sequences. 
Additionally, there is no statistical error present for the self-inverting circuit under this error model because all circuits of the same sequence length have exactly the same overall error, as the error channel commutes with all the gates in $Cl_1$.
In contrast, for a gate-independent unitary error, the scaling remains linear for the self-inverting circuits but exhibits a $\sqrt{m}$ scaling for generic circuits.  
This occurs because when the state system is in the xy-plane before measurement, each error contributes a random sign to the probability of each outcome, whereas when the system is on the $z$ axis each error has to contribute a systematic sign \cite{wallman2015bounding}. 
As shown in \cref{New-metric-plot}, this implies that restricting to self-inverting circuits can underestimate the MVE by over an order of magnitude in the small-$m$ regime, which is relevant for near-term quantum computer applications.

Unlike other distance measures where an improvement in quality can be caused by a bias in choosing a gauge, a decrease in MVE is unequivocally an improvement due to its gauge-invariance and because, by definition, the output probability distribution gets closer to the ideal distribution.
Furthermore, the MVE captures the relevant behavior for generic circuits, rather than just self-inverting circuits which, by design, perform no useful computation.
  
A protocol for estimating the MVE of a gate-set $\Phi = \{\mathds{S}, \mathds{G}, \mathds{M}\}$ is as follows:
\begin{enumerate}
	\item Select $N_m$ random experiments $C\in\mathds{A}_m$, for some $N_m$ large enough to accurately approximate the average.
	\item Repeat each experiment $C$ $K_m$ times to estimate $\langle \tilde{M}_i, \tilde{G}_{m:1}(\tilde{\rho}) \rangle$ for each $C$. 
	\item Compute the ideal probabilities $\langle M_i, G_{m:1}(\rho) \rangle$ for each observed outcome of $\tilde{C}$. 
	\item Calculate $\delta d(C, \tilde{C})$ for each experiment $C$, average over them to estimate $x(\Phi, m)$.
	\item Repeat step 1--4 for different values of $m$ to measure the scaling behaviour of MVE.
\end{enumerate}
Note that if $\mathds{G}$ is a unitary 2-design and the states and measurements are chosen appropriately, the applied operations are identical to those used to estimate the unitarity \cite{Wallman2015}.
The primary difference in the protocol is that it is more scalable, more general, and has different post-processing.

The scalability of the above protocol is affected by the number of experiments $N_m$, the number of repetitions for each experiment $K_m$, and the complexity of calculating the ideal probabilities.
The number of experiments determine the accuracy of the MVE, and can be estimated using Hoeffding's inequality independently of the number of qubits \cite{hoeffding1963probability}.
The complexity in the protocol is determined by the complexity of calculating probabilities and by the number of repetitions required to estimate $\delta d(C,\tilde{C})$ to a fixed precision.
The number of repetitions required to estimate $\delta d(C,\tilde{C})$ to a fixed precision is polynomial in the number of outcomes \cite{chan2014}.
To efficiently characterize multi-qubit gate-sets (where the number of raw outcomes grows exponentially with the number of qubits), we can coarse-grain the measurements over sets of outcomes.
The computational complexity of calculating each probability will depend on the gate-set in question.
The ideal probabilities can be efficiently computed if $\mathds{G}$ is the $N$-qubit Clifford group \cite{gottesman1998heisenberg,nest2008classical}.
For gate-sets containing only one- and two-qubit gates and product states and measurements, the MVE can be computed for small values of $m$.
However, for a generic gate-set, each probability will be hard to compute.
Of course, for small systems with a few qubits, this procedure can nonetheless be performed quickly on a classical computer.

An experimentalist can perform a feedback loop whereby they update the control parameters, rerun the MVE evaluation experiment (potentially for some fixed value of $m$) and compare to the previous result to see if the error has decreased.
Protocols that use feedback from experimental outcomes to improve control over quantum devices have been proposed before, such as in \cite{kelly2014optimal} where control parameters were optimized by maximizing the randomized benchmarking survival probability for a fixed sequence length.
Optimizing the MVE instead of the randomized benchmarking survival probability corresponds to minimizing the effect of errors on generic quantum circuits, rather than minimizing the effect of errors on self-inverting circuits.
As demonstrated in \cref{New-metric-plot}, errors in self-inverting circuits may be substantially smaller than those in generic circuits because such circuits suppress coherent errors and implement a form of randomized dynamical decoupling \cite{viola2005}.

\begin{acknowledgments}
JJW acknowledges helpful discussions with Robin Blume-Kohout.
This research was supported by the U.S. Army Research Office through grant W911NF-14-1-0103, the Government of Ontario, and the Government of Canada through CFREF, NSERC and Industry Canada.
\end{acknowledgments}

\clearpage
\newpage
\bibliography{Main}

\appendix
\section{Pauli-Liouville representation}\label{appendix}

Here we present the Pauli-Liouville representation for gate-set elements.
For simplicity we focus on system of $n$ qubits where quantum states can be represented as $2^{n} \times 2^{n}$ Hermitian operators, while the same formalism can be generalized to arbitrary dimension systems by using, for example, the generalized Gell-Mann matrices~\cite{bertlmann2008bloch} instead of Pauli matrices.
It is commonly known that the set of (tensor products of) normalized Pauli matrices, which we denote as $\mathbf{P}_n$, form an orthonormal basis for all $2^{n} \times 2^{n}$ Hermitian operators.
Every element $P$ in $\mathbf{P}_n$ is of the form
\begin{equation}
P = \bigotimes_{k} \left(\frac{\sigma_k}{\sqrt{2}}\right)
\end{equation}
and each $\sigma_k$ is a member from the single-qubit Pauli group, $\mathbf{P}_1 \coloneqq \{\sigma_0 = I,\ \sigma_1 = X,\ \sigma_2 = Y,\ \sigma_3 = Z\}$. 
The orthonormality is defined with respect to the Hilbert-Schmidt inner product
\begin{equation}
\langle P_i,\ P_j \rangle_{HS} \coloneqq \Tr[P_i^\dagger P_j] = \delta_{ij}
\end{equation}
for all $P_i, P_j \in \mathbf{P}_n$.
Any $2^n$ by $2^n$ Hermitian matrix can be represented as a real linear combination of Pauli basis matrices.
Writing these inner products as components of a vector will define a representation in the space of $2^{2n} \times 1$ real vectors, which is isomorphic to the set of $2^n \times 2^n$ real matrices.

For every $2^n$ by $2^n$ Hermitian matrix $\rho$, we define its Pauli-Liouville representation as follows:
\begin{equation}
\dket{\rho} \coloneqq \sum_{i} \Tr[\rho P_i] \dket{i}
\end{equation}
and define an element $\sigma$ in the dual space (e.g., representing a measurement operator) as
\begin{equation}
\dbra{\sigma} \coloneqq \sum_{i} (\Tr[P_i \sigma])^* \dbra{i}
\end{equation}
where $\dket{i}$ and $\dbra{i}$ are standard computational (column and row) basis vectors with $1$ in the $i$-th entry and $0$ elsewhere, and $^*$ denotes complex conjugation for a scalar.
We see that the Hilbert-Schmidt inner product is now transformed into an Euclidean inner product:
\begin{equation}
\begin{split}
\dbraket{\sigma}{\rho} &= \sum_{i,j} \Tr[\rho P_i] (\Tr[ P_j \sigma])^* \dbraket{j}{i}\\
&= \sum_{i,j} \Tr[\rho P_i] (\Tr[P_j \sigma])^* \Tr[P_i P_j]\\
&= \Tr[\sum_{i} \Tr[\rho P_i] P_i \sum_j (\Tr[P_j \sigma])^* P_j]\\
&= \Tr[\sigma^\dagger \rho]
\end{split}
\end{equation}
where the following identity is used:
\begin{equation}
\begin{split}
\sum_i \Tr[P_i A]^* P_i & = \sum_i \left(\Tr[P_i^\dagger A]\right)^* P_i\\
&= \sum_i \Tr[\overline{(A^\dagger P_i)^\dagger}] P_i\\
&= \sum_i \Tr[(A^\dagger P_i)^T] P_i\\
&= \sum_i \Tr[A^\dagger P_i] P_i = A^\dagger
\end{split}
\end{equation}
where the overhead bar denotes element-wise conjugation for a matrix.

Now, define the Pauli-Liouville representation of a (linear) map $\mc{G}$ as $\mc{A}_{\mc{G}}$, which has components
\begin{equation}
(\mc{A}_{\mc{G}})_{i j} \coloneqq  \Tr[P_{i} \mc{G}(P_{j})] 
\end{equation}
then the post-state of $\mc{G}$ acting on a state $\rho$, written all in the Pauli basis, can be shown to be equal to a matrix multiplication:
\begin{equation}\label{PauliProduct}
\begin{split}
\dket{\mc{G}(\rho)} & = \sum_{i} \Tr[\mc{G}(\rho) P_{i}] \dket{i}\\
&= \sum_{i} \Tr[\mc{G}(\sum_{j} \Tr[\rho P_{j}] P_{j}) P_{i}] \dket{i}\\
&= \sum_{i j} \Tr[\rho P_{j}] \Tr[\mc{G}( P_{j}) P_{i}] \dket{i}\\
&= \sum_{i j} (\mc{A}_{\mc{G}})_{i j} (\dket{\rho})_{j} \dket{i}\\
&= \mc{A}_{\mc{G}} \dket{\rho}
\end{split}
\end{equation}
Thus, series of gates are conveniently expressed as matrix multiplications (from the left) in the Pauli-Liouville representation.

\end{document}